\documentclass[a4paper,11pt]{article}
\pdfoutput=1 

\usepackage{jheppub} 

\usepackage[T1]{fontenc} 
\usepackage{amsmath,amsfonts,epsfig,color,latexsym}
\usepackage{amssymb}
\usepackage[USenglish]{babel}
\usepackage{epsfig,psfrag}
\usepackage{rotating}
\usepackage{hyperref}
\usepackage{pdfpages}
\usepackage{multirow}

\usepackage{xcolor}

\preprint{MITP/18-077 \\
	\phantom{~} \hfill UUITP-34/18 \\
	\phantom{~} \hfill TCDMATH 18-09}

\makeatletter
\renewcommand\@fpheader{} 
\renewcommand\@journal{}
\makeatother

\title{\boldmath All five-loop planar four-point functions of half-BPS operators in $\mathcal N=4$ SYM}

\author[\dagger]{Dmitry Chicherin,}
\author[\mathcal{x}]{Alessandro Georgoudis,}
\author[\mathcal{w}]{Vasco Gon\c calves}
\author[\mathcal{z}]{and Raul Pereira}

\affiliation[\dagger]{PRISMA Cluster of Excellence, Johannes Gutenberg University, 55099 Mainz, Germany}
\affiliation[\mathcal{x}]{Department of Physics and Astronomy, Uppsala University
Box 516, SE-751 20 Uppsala, Sweden}
\affiliation[\mathcal{w}]{ICTP South American Institute for Fundamental Research, IFT-UNESP,\\
S\~ao Paulo, SP Brazil 01440-070}
\affiliation[\mathcal{z}]{School of Mathematics and Hamilton Mathematics Institute, Trinity College Dublin, \\Dublin, Ireland}

\newcommand{\p}[1]{(\ref{#1})}

\newcommand \vev [1] {\langle{#1}\rangle}

\newcommand{\cL}{{\cal L}} 

\newcommand{\cN}{{\cal N}}

\newcommand{\cO}{{\cal O}}

\newcommand{\la}{\lambda}

\abstract{We obtain all planar four-point correlators of half-BPS operators in $\mathcal{N}=4$ SYM up to five loops. The ansatz for the integrand is fixed partially by imposing light-cone OPE relations between different correlators. We then fix the integrated correlators by comparing their asymptotic expansions with simple data obtained from integrability. We extract OPE coefficients and find a prediction for the triple wrapping correction of the hexagon form factors, which contributes already at the five-loop order.}

\begin{document}
\maketitle
\flushbottom




\section{Introduction}

Correlation functions of local operators are among the most interesting observables to be studied in a CFT. They encode nontrivial physics of the theory that can be accessed using different limits of the correlation functions (large spin, bulk point or Regge limit \cite{Alday:2013cwa,Maldacena:2015iua,Costa:2012cb}). Of all CFTs  known, $\mathcal{N}=4$ SYM stands at a special point where symmetries of the theory might allow to completely solve it. It is then possible to study the effects of finite coupling in a four-dimensional gauge theory, which might lead to better strategies in the study of other quantum field theories.  

The most powerful method in $\mathcal{N}=4$ SYM that exploits these symmetries is integrability, which started with the understanding of two-point functions of single-trace operators in the planar limit \cite{Gromov:2013pga,Gromov:2014caa,Gromov:2015wca}. More recently it was understood how to use integrability to compute higher-point correlators of local operators \cite{BKV,Fleury:2016ykk,Fleury:2017eph,Eden:2016xvg} and even to obtain non-planar quantities \cite{Eden:2017ozn,Bargheer:2017nne}. This proposal, known as the hexagon approach, has now passed many non-trivial checks both at weak and strong coupling \cite{Eden:2015ija,3loops,Basso:2017muf,Vasco,EdenPaul,Eden,JT}. However, despite being a finite-coupling proposal this program is taking its first steps and there are still aspects that need to be better understood, so it is essential to obtain field-theoretic results which provide further checks and clarify subtleties within the integrability framework.

Correlators of half-BPS scalar operators are probably the simplest objects in $\mathcal{N}=4$ SYM, and the fact that they are finite and do not need infinite renormalization makes them ideal objects to study. While two- and three-point functions are protected, higher-point functions have an explicit coupling dependence, which motivated their study in the early days of AdS/CFT correspondence, both at weak and strong coupling \cite{Eden:2000mv,Arutyunov:2000py,Arutyunov:2002fh,Arutyunov:2018tvn}. More recently, the discovery of a symmetry enhancement \cite{Eden:2011we} has been combined with a light-cone OPE analysis, which allowed to fix the correlator of four $\mathcal{O}_{20'}$ operators to very high loop order \cite{Bourjaily:2016evz}. This OPE constraint is very powerful, as it implies exponentiation of the correlator in the light-cone limit, therefore providing recursive relations between different orders in the perturbative expansion of the four-point function.  Let us remark that some correlators have also been obtained using bootstrap methods \cite{Rastelli:2016nze,Rastelli:2017udc,Rastelli:2017ymc,Alday:2017xua,Aprile:2017xsp,Aprile:2017qoy,Goncalves:2014ffa}.

The goal of this paper is to compute the four-point correlation functions of half-BPS operators with higher $R$-charge weights, up to five loops. In these generic configurations the symmetry mentioned above is not as strong and the light-cone OPE not as constraining, which means that the integrand cannot be completely determined with these methods. In this work we combine the light-cone OPE analysis with OPE data extracted from integrability, and successfully fix all four-point functions at four and five loops. We want to emphasize that we only needed OPE coefficients that are quite easy to obtain from the integrability point of view, while the data extracted from the four-point functions allows us to make highly non-trivial predictions for finite-size corrections of hexagon form factors. The most important result is the leading five-loop order of the triple wrapping correction, which was originally expected to contribute only from six loops.

In Section \ref{4ptcorrelators} we describe the symmetries of the correlator's integrand, which allow us to construct an ansatz given in terms of conformal integrals. In Section \ref{secLCOPE} we show how to fix most coefficients in the ansatz by relating the light-cone OPE limit of correlators with different weights. We follow with Section \ref{condInt} where we explain how one can use input from integrability to fix the remaining coefficients. We then present our results for the correlators at four and five loops in section \ref{res_sec}, where we also elaborate on the predictions for finite-size correction of hexagon form factors that we can extract from the euclidean OPE limit of the four-point functions. We end in Section \ref{conc} with our conclusions and future research directions.  Finally,  appendix \ref{AppAE} contains a short review on asymptotic expansions of conformal integrals. We also provide an auxiliary file with all four- and five-loop four-point functions, as well as the leading asymptotic expansions for all relevant integrals at that loop order.

\section{Four-point correlation functions and integrands}\label{4ptcorrelators}
We consider gauge-invariant operators at the bottom of half-BPS supermultiplets of $\cN = 4$ SYM theory.
The operator of weight $L$ is realized as a single trace of the product of $L \geq 2$ fundamental scalars $\Phi^I(x)$, $I =1,\ldots,6$,
\begin{align} \label{ophalfBPS}
\cO_L(x,y) = y_{I_1} \ldots y_{I_L} {\rm Tr} \left( \Phi^{I_1} \ldots \Phi^{I_L} \right)(x)\,. 
\end{align}
The traceless symmetrization over R-symmetry indices is provided by the auxiliary $\mathfrak{so}(6)$ harmonic variables $y_I$: $y \cdot y = 0$.
Half-BPS operators are protected -- they do not undergo infinite renormalization, so their conformal dimension exactly equals to $L$ and the correlation functions of these operators are finite quantities in $D=4$. Also the classical (super)conformal symmetry of the ${\cal N} = 4$ SYM Lagrangian is inherited by these dynamical quantities. The two- and three-point correlation functions are completely fixed by the conformal symmetry, and their tree-level approximation is exact. For more points the correlators receive quantum corrections.
We study the four-point correlators
\begin{align} \label{4pnt}
\vev{\cO_{L_1}(x_1,y_1) \cO_{L_2}(x_2,y_2) \cO_{L_3}(x_3,y_3) \cO_{L_4}(x_4,y_4)}\,.
\end{align}
They are highly nontrivial functions containing useful information about dynamics of the theory. 
At the same time the symmetry constraints considerably simplify their form that makes them more manageable as compared with higher-point correlators.

In the tree approximation the correlators are given by the sum of products of free propagators $d_{ij} = \frac{y^2_{ij}}{x_{ij}^2}$ stretched between scalar fields $\Phi$. Here $y_{ij}^2 \equiv y_i \cdot y_j$ and $x_{ij}^2 \equiv (x_i - x_j)^2$. 

The perturbative expansion of the correlators in the `t Hooft coupling $\la=g^2 N_c/ (4 \pi^2)$ 
contains a huge number of Feynman diagrams which have to be added together to obtain a gauge-invariant quantity. Thus, prior to any loop integrations, just finding the gauge-invariant integrand of correlator \p{4pnt} constitutes a nontrivial problem. In this paper we solve this problem up to the five-loop order for arbitrary BPS weights using the integrability methods.

The Lagrangian insertion formula \cite{Eden:2000mv} provides a neat expression for the integrand of \p{4pnt}
\begin{align}
&\vev{\cO_{L_1} \cO_{L_2} \cO_{L_3} \cO_{L_4}}_{(\ell)} \sim 
\int d^4 x_5 \ldots d^4 x_{4+\ell}\, 
\vev{\cO_{L_1} \ldots \cO_{L_4} \cL(x_5)\ldots \cL(x_{4+\ell})}_{\text{Born}}
\,,  \label{integrand}
\end{align} 
as the correlation function of $4+\ell$ operators -- four operators $\cO_{L_i}$ and $\ell$ chiral Lagrangian densities $\cL$ -- calculated in the Born approximation, which is the lowest nontrivial perturbative approximation. Let us stress that the Born level $(4+\ell)$-point correlator 
\begin{align}
G^{(\ell)}_{L_1,L_2,L_3,L_4} \equiv \vev{\cO_{L_1}(x_1,y_1) \ldots \cO_{L_4}(x_4,y_4) \cL(x_5) \ldots \cL(x_{4+\ell})}_{\rm Born} \label{born}
\end{align} 
is of order $\la^\ell$, and familiar Feynman diagrams representing this correlator involve the interaction vertices. Nevertheless, $G$ is a rational function of $4+\ell$ space-time coordinates $x$ and it is polynomial in harmonic variables $y$. $G$ carries conformal weight $L_i$ and harmonic weight $L_i$ at external points ${\cal E} = \{1,2,3,4\}$, and 
zero harmonic weight and conformal weight $(+4)$ at internal points ${\cal I} = \{5,\ldots,4+\ell\}$. 
$G$ is a particular component of the supercorrelator of $4+\ell$ half-BPS multiplets. The super-conformal symmetry of the latter implies \cite{Eden:2000bk,Heslop:2002hp,Nirschl:2004pa,Eden:2011we} that $G$ is proportional to the rational factor $R(1,2,3,4)$,
\begin{align}\label{R4}
R(1,2,3,4) & =  d_{12}^2 d_{34}^2 x_{12}^2 x_{34}^2 +  d_{13}^2 d_{24}^2 x_{13}^2 x_{24}^2 +  d_{14}^2 d_{23}^2 x_{14}^2 x_{23}^2
\notag\\ 
& +d_{12} d_{23} d_{34} d_{14} ( x_{13}^2 x_{24}^2 -x_{12}^2 x_{34}^2 - x_{14}^2 x_{23}^2) \notag \\
& +d_{12} d_{13} d_{24} d_{34} ( x_{14}^2 x_{23}^2 -x_{12}^2 x_{34}^2 - x_{13}^2 x_{24}^2) \notag \\
& +d_{13} d_{14} d_{23} d_{24} ( x_{12}^2 x_{34}^2 -x_{14}^2 x_{23}^2 - x_{13}^2 x_{24}^2)\,.
\end{align}
This factor absorbs harmonic weight $(+2)$ and conformal weight $(+1)$ at external points ${\cal E}$.
The complementary harmonic weights, i.e. $L_i-2$ at point $i \in {\cal E}$, can be absorbed by propagator factors, that leads to the following generic form of the Born-level correlator 
\begin{align}\label{integrand2}
G^{(\ell)}_{L_1,L_2,L_3,L_4} = \la^{\ell}\, C_{L_1 L_2 L_3 L_4}\, R(1,2,3,4)\, 
\sum_{\{b_{ij}\}}\left(\prod_{\substack{i < j\\ i,j \in {\cal E}}} (d_{ij})^{b_{ij}} \right) 
\frac{P^{(\ell)}_{\{b_{ij}\}}(x_1,\ldots,x_{4+\ell})}{\prod\limits_{\substack{p \in {\cal E} \\ q \in {\cal I}}} x_{pq}^2 \prod\limits_{\substack{p<q \\ p,q \in {\cal I}}} x_{pq}^2} \,.
\end{align}
The summation in eq. \p{integrand2} is over tuples $\{b_{ij}\}^{i<j}_{i,j\in{\cal E}}$ satisfying constraints $\sum_{j\neq i} b_{ij} = L_i-2$ for each $i \in {\cal E}$. The tuples represent different ways to distribute harmonic weights. Then the conformal weight counting shows that $P^{(\ell)}_{\{b_{ij}\}}$ carries weight $(1-\ell)$ at each point ${\cal E}\cup {\cal I}$.
The numerical normalization factor $C$ in \p{integrand2} is chosen for the sake of convenience, 
\begin{align}\label{}
C_{L_1 L_2 L_3 L_4} =  \frac{L_1 L_2 L_3 L_4}{2(4\pi^2)^{\frac{1}{2}\sum L_i}} \left(\frac{N_c}{2}\right)^{\frac12 \sum L_i -2}\,. \label{defC}
\end{align} 
A simple short-distance OPE analysis reveals that $G \sim 1/x_{pq}^2 + O(1)$  at $x_p \to x_q$ if $p\in {\cal E}$ and $q \in {\cal I}$ or $p,q \in {\cal I}$. This implies that 
$P^{(\ell)}_{\{b_{ij}\}}$ in eq. \p{integrand2} is polynomial in space-time coordinates.
The polynomial $P^{(\ell)}_{\{b_{ij}\}}$ has certain discrete symmetries. E.g. the integrand of the four-point function of ${\cal O}_{\bf 20'}$ operators ($L_1 = \ldots = L_4 =2$) is specified by one conformal polynomial with $\{b_{ij}\} = \{ 0,0,0,0,0,0\}$ which is invariant under all permutations ${\cal S}_{4+\ell}$ of $(4+\ell)$ space-time points \cite{Eden:2011we}.
In the case of generic half-BPS weights the conformal polynomial $P^{(\ell)}_{\{b_{ij}\}}$ has the reduced discrete symmetry. It is invariant with respect to the same subgroup $\mathfrak{G} \subset {\cal S}_{4+\ell}$, acting on points ${\cal E}\cup {\cal I}$, as the accompanying factor 
\begin{align}
\prod_{\substack{i < j \\  i,j \in {\cal E} }} (d_{ij})^{b_{ij}}\,.
\end{align} 
Obviously $\mathfrak{G}$  contains ${\cal S}_{\ell}$ as a subgroup, ${\cal S}_{\ell} \subset \mathfrak{G}$ , which acts on the Lagrangian points.

Thus the construction of the correlator integrand boils down to fixing a number of conformal polynomials $P^{(\ell)}_{\{b_{ij}\}}$ 
with given discrete symmetries.
There is a finite number of them at each loop order $\ell$ and they can be enumerated. Therefore the remaining freedom reduces to a number of numerical constants.

Integrating out $\ell$ internal points ${\cal I}$ according to \p{integrand} we rewrite the contribution of each $SU(4)$ harmonic structure
in \p{integrand2} as a linear combination of $\ell$-loop four-point conformally covariant integrals $I^{(\ell)}(1,2,3,4)$,
\begin{align}
\int d^4 x_1 \ldots d^4 x_{4+\ell}\frac{P^{(\ell)}_{\{b_{ij}\}}(x_1,\ldots,x_{4+\ell})}{\prod\limits_{\substack{p \in {\cal E} \\ q \in {\cal I}}} x_{pq}^2 \prod\limits_{\substack{p<q \\ p,q \in {\cal I}}} x_{pq}^2} = \sum_k c^{(k)}_{\{b_{ij}\}} \, I^{(\ell)}_{k}(1,2,3,4) \label{Pint}
\end{align}
where the numerical coefficients $c^{(k)}_{\{b_{ij}\}}$ are the same as in monomials of the conformal polynomials $P^{(\ell)}_{\{b_{ij}\}}$. An integral $I(1,2,3,4)$ carries weights $(+1)$ at all four external points, so it can be represented as 
\begin{align} \label{IIuv}
I(1,2,3,4) = \frac{1}{x_{13}^2 x_{24}^2}\, I(u,v)
\end{align}
where $I(u,v)$ is a conformally invariant function and, consequently, it depends on conformal cross-ratios
\begin{align}\label{cross}
u=\frac{x^2_{12}\, x^2_{34}}{x^2_{13}\, x^2_{24}}\;,\;\;\;
v=\frac{x^2_{14}\, x^2_{23}}{x^2_{13}\, x^2_{24}}\,.
\end{align}
Several examples of five-loop conformally covariant integrals are given in eq.~\p{confintex}.

The number of linear independent conformal integrals is smaller than one could naively expect on the basis of the discrete symmetries of their integrands. The conformal symmetry implies nontrivial relations among them, e.g. \begin{align} \label{IeqI} 
I(1,2,3,4)= I(3,4,1,2)
\end{align} 
immediately follows from \p{IIuv}. The latter relation reduces the number of independent orientations of the given integral. 
Applying \p{IeqI} to the conformal $\ell'$-loop subintegrals ($\ell' < \ell$) of the $\ell$-loop integrals we generate 'magic' identities \cite{Drummond:2006rz} among $\ell$-loop integrals of the different topology. Also some of the $\ell$-loop integrals trivially factorize in a product of several lower-loop conformal integrals, and some of the integrals differ only by a rational factor in cross-ratios $u,v$. These observations enable us to reduce the number of conformal integrals we have to deal with. The number of non-trivially distinct $\ell$-loop integrals is given in Tab.~\ref{tab_confint}. The asymptotic expansion of the integrals at $u \to 0, v\to 1$ is discussed in App.~\ref{AppAE} and the results are collected in an ancillary file. 

\begin{table}
\begin{center}
\begin{tabular}{c||c|c|c|c|c}
loop order $\ell$ & 1 & 2 & 3 & 4 & 5 \\ \hline
$\#$ of integrals & 1 & 1 & 3 & 19 & 141
\end{tabular}
\end{center}
\caption{The number of $\ell$-loop integrals $I^{(\ell)}(u,v)$ contributing to the correlators \p{confint}. For the sake of simipicity we mode out: 1). integrals, which factorize in a product of lower-loop integrals; 2). permutations of external points; 3). rational factors in cross-ratios $u,v$ accompanying conformal integrals.} \label{tab_confint}
\end{table}

In the following we denote \p{Pint} -- the integrated contribution of the $\{b_{ij}\}$ harmonic structure to the rhs of eq.~\p{integrand2} -- by $\frac{F^{(\ell)}_{\{b_{ij}\}}(u,v)}{x^2_{13} x^2_{24}}$. As we discussed above it is given by a linear combination of the conformal integrals
\begin{align} \label{confint}
F^{(\ell)}_{\{b_{ij}\}}(u,v) = \sum_{m} \tilde c^{(m)}_{\{b_{ij}\}} \, I^{(\ell)}_{m}(u,v)\,,
\end{align}
where numerical coefficients $\tilde c^{(m)}_{\{b_{ij}\}}$ are linear combinations of $c^{(k)}_{\{b_{ij}\}}$ originating from conformal polynomials 
$P^{(\ell)}_{\{b_{ij}\}}$. Let us stress that the integrated expression \p{confint} contains less coefficients than the integrand.
Thus we obtain 
the following representation for the four-point correlator
\begin{align}\label{loopCorr}
\vev{\cO_{L_1} \cO_{L_2} \cO_{L_3} \cO_{L_4}}_{(\ell)} = \la^{\ell}\, C_{L_1 L_2 L_3 L_4}\, R(1,2,3,4)\, 
\sum_{\{b_{ij}\}}\left(\prod_{\substack{i < j\\ i,j \in {\cal E}}} (d_{ij})^{b_{ij}} \right) \frac{F^{(\ell)}_{\{b_{ij}\}}(u,v)}{x^2_{13} x^2_{24}}\,.
\end{align}

The correlator is specified by weights $\{ L_i \}_{i \in {\cal E}}$ of the half-BPS operators, and correlators of different weights do not have to coincide. However in each given loop order $\ell$ there is only a finite number of different correlators. This is rather obvious from the point of view of Feynman graphs. Indeed, there is no more than $2\ell$ interaction vertices in the corresponding Feynman graphs, consequently for sufficiently large weights $\{L_i\}$ some propagators are spectators. They are stretched between pairs of operators $\cO_{L_i}$ and $\cO_{L_j}$ like in tree graphs. Thus there is a finite number of functions $F^{(\ell)}_{\{b_{ij}\}}$ at any given loop order $\ell$. More precisely, there is a saturation bound $\kappa=\kappa(\ell)$ such that 
\begin{align}\label{bound}
F^{(\ell)}_{\{ b_{12},b_{13},b_{14},b_{23},b_{24},b_{34} \} } = F^{(\ell)}_{ \{ \kappa,b_{13},b_{14},b_{23},b_{24},b_{34} \} } \qquad \text{at}\;\; b_{12} \geq \kappa,\;  b_{13},\ldots, b_{34} \geq 0 \,
\end{align}     
and similar relations also hold for any other index $b_{ij}$ instead of $b_{12}$. We expect that minimal value of the saturation bound is
\begin{align}\label{bound2}
\kappa_{\rm min}(\ell) \equiv {\rm min}\; \kappa(\ell) = \ell - 1 \,.
\end{align}
Previously it has been proven to be true up to the three-loop order. We argue that it should hold up to the five-loop order. Choosing the saturation bound $\kappa$ in \p{bound} higher than $\kappa_{\rm min}$ and implementing the correlator bootstrap we find that relations \p{bound} hold with $\kappa=\kappa_{\rm min}$.
In Tab.~\ref{num_fun} we show the number of  functions $F^{(\ell)}_{\{b_{ij}\}}$  for $\kappa = \kappa_{\rm min}$ modding out permutations of the external points.

\begin{table}
\begin{center}
\begin{tabular}{c||c|c|c|c|c}
loop order $\ell$ & 1 & 2 & 3 & 4 & 5 \\ \hline
$\#$ of $\{b_{ij}\}$ & 1 & 11 & 66 & 276 & 900
\end{tabular}
\end{center}
\caption{The number of different harmonic structures (modulo permutation of the external points) specified by $\{ b_{ij}\}$ in the set of all $\ell$-loop correlators assuming that the saturation bound in \p{bound} is $\kappa = \kappa_{\rm min}$.} \label{num_fun}
\end{table}

\section{Correlator bootstrap with light-cone OPE}\label{secLCOPE}

Up to now we have not used planarity restrictions. In order to make use of some dynamical constraints on coefficients of polynomials $P^{(\ell)}_{\{b_{ij}\}}$ we consider the planar approximation. In particular we imply that the graphs representing the integrand $G$, eq. \p{integrand2}, have planar topology. In this way we considerably reduce the number of admissible polynomials $P^{(\ell)}_{\{b_{ij}\}}$. Then we can try to fix the remaining numerical coefficients by means of the OPE analysis. 

We would like to impose OPE constraints directly on the integrands. Obviously it is preferable to deal with the rational integrands than with unknown multi-loop integrals. In this way we try to pin down as many coefficients in the ansatz \p{integrand2} as possible. Then we fix the remaining coefficients by extracting more detailed dynamical information from the OPEs of the integrated quantities.   

In \cite{Eden:2012tu} the four-point correlator $\vev{{\cal O}_{\bf 20'} {\cal O}_{\bf 20'} {\cal O}_{\bf 20'} {\cal O}_{\bf 20'}}$ of weights $L_1 = L_2 = L_3 =L_4 = 2$ was considered, and constraints on the asymptotic behavior of its integrand were found in the light-cone limit $x_{12}^2,x_{23}^2,x_{34}^2,x_{14}^2 \to 0$. The correlator exponentiates in this limit that implies relations among different orders of the perturbative expansion, so the correlator can be recursively constrained order by order. Using this approach the integrands have been fixed up to three loops at generic $N_c$ \cite{Eden:2012tu} and up to ten loops in the planar limit \cite{Eden:2012tu,Ambrosio:2013pba,Bourjaily:2015bpz,Bourjaily:2016evz}. 

For higher-weight correlators a similar exponentiation property does not seem to hold. Nonetheless some useful OPE constraints for the integrands are known. In \cite{Chicherin:2015edu} studying the light-cone OPE $x_{12}^2 \to 0$ of higher-weight Born-level correlators \p{born} in the planar approximation
the following relation was obtained
\begin{align}\label{LCOPE}
 \lim_{\substack{x_{12}^2\rightarrow 0 \\ y_1 \to y_2 \\
  d_{12} \text{ fixed}}}  \, \left[ \frac{G^{(\ell)}_{L_1+1, L_2+1, L_3 L_4}}{C_{L_1+1, L_2+1, L_3 L_4}}  
	- d_{12}\times \frac{G^{(\ell)}_{L_1 L_2 L_3 L_4}}{C_{L_1 L_2 L_3 L_4}} \right] = O(d_{12}) \,
\end{align}
where $C$ is defined in \p{defC}.
It compares the leading light-cone singularities of a pair of integrands with different BPS weights. Using \p{LCOPE} the correlator integrands of all weights have been fixed up to the three-loop order in the planar approximation.  

Let us briefly explain the origin of eq.~\p{LCOPE} following \cite{Chicherin:2015edu}. We consider the contribution of a non-protected operator ${\cal O}_{L,S}$ of twist $L$, spin $S$, which belongs to some representation of $SU(4)$, in the OPE of two half-BPS operators  at $x_{12}^2 \to 0$, i.e. schematically ${\cal O}_{L_1} \times {\cal O}_{L_2} \to {\cal O}_{L,S}$. This contribution is proportional to the structure constant $C_{L_1, L_2, {\cal O}_{L,S}}(\lambda) \sim \vev{{\cal O}_{L_1} {\cal O}_{L_2} {\cal O}_{L,S}}$, so inserting it in the Born-level correlator \p{born} we obtain
$G^{(\ell)}_{L_1,L_2,L_3,L_4} \sim C_{L_1, L_2, {\cal O}_{L,S}} \vev{{\cal O}_{L,S} {\cal L} \ldots {\cal L}} $ at $x_{12}^2 \to 0$. The tree-level structure constants in the planar approximation satisfy the following relation 
\begin{align}
\frac{C_{L_1+1,L_2+1,{\cal O}_{L,S}}}{C_{L_1,L_2,{\cal O}_{L,S}}} = \frac{C_{L_1+1,L_2+1,L_3,L_4}}{C_{L_1,L_2,L_3,L_4}} \,.
\end{align}
Consequently, if we could use the tree-level approximation for $C_{L_1, L_2, {\cal O}_{L,S}}$ then the OPE contribution of ${\cal O}_{L,S}$ cancels in the difference of correlators $G^{(\ell)}_{L_1+1, L_2+1, L_3 L_4}$ and $G^{(\ell)}_{L_1, L_2, L_3 L_4}$ from eq. \p{LCOPE}. In particular it is true for the operators from $\mathfrak{sl}(2)$ sector (see Sect. \ref{sl2}). In order to isolate the appropriate OPE channels we take the limit in \p{LCOPE}. 
If we could use the tree-level approximation for the structure constants of generic operators ${\cal O}_{L,S}$ then  a stronger version of \p{LCOPE} should hold
\begin{align}\label{strong}
 \frac{G^{(\ell)}_{L_1+1, L_2+1, L_3 L_4}}{C_{L_1+1, L_2+1, L_3 L_4}}  
	- d_{12}\times \frac{G^{(\ell)}_{L_1 L_2 L_3 L_4}}{C_{L_1 L_2 L_3
  L_4}} = O\left(\frac{1}{x_{12}^2}\right)\qquad \text{at} \;\; x_{12}^2 \sim 0 \,,
\end{align}
which was conjectured in~\cite{Chicherin:2015edu}.
At $\ell \leq 3$ loop order it is equivalent to \p{LCOPE}, but starting from four loops \p{strong} is more restrictive. Let us remark that the strong criterion implies the saturation bound $\kappa=\kappa_{\rm \min}$ \p{bound2} at least up to five loops.

We are going to constraint all higher-weight correlators at four- and five-loops in the planar approximation. 
For the bootstrap procedure it is essential to consider correlators of all weights simultaneously rather than their subset, since relations \p{LCOPE} are more restrictive in the former case. We use the weight-two correlator integrands $G^{(\ell)}_{2,2,2,2}$ from \cite{Eden:2012tu} as an input and constrain higher-weight correlators. Also we make use of additional constraints on the integrand $G^{(\ell)}_{3,3,2,2}$ following from exponentiation property of the short-distance OPE $x_1 \to x_3$ \cite{Chicherin:2015edu,Eden:2012tu} for the corresponding four-point correlator.
Neither weak \p{LCOPE} nor strong \p{strong} criteria are enough to fix all coefficients starting from the four-loop order. Nevertheless, they considerably reduce the number of unknowns, see Tab.~\ref{tab_ws}. In the following we apply the weak criterion to partially fix the integrand and then we use integrability of the three-point functions to pin down the remaining coefficients. The obtained results are in agreement with the strong criterion \p{strong}.

\begin{table}
\begin{center}
\begin{tabular}{|c|c||c|c|c||c|}
loop order $\ell$ & bound $\kappa$ & planar + sym & weak & strong & OPE $\vev{3322}$ \\
\hline 1 & 0 & 0 & & &\\
\hline 2 & 1 & 14 & 0 & 0 &\\
\hline 3 & 2 & 347 & 1 & 1 & -1 \\
\hline \multirow{3}{*}{4} & 3 & 8543 & 37 & \multirow{3}{*}{6} & \multirow{3}{*}{-3}\\
\cline{2-4}  & 4 & 24749 & 77 &  & \\
\cline{2-4}  & 5 & 59234 & 149 &  & \\
\hline \multirow{2}{*}{5} & 4 & 191372 & 614 & \multirow{2}{*}{33} & \multirow{2}{*}{-12} \\  
\cline{2-4} & 5 & 459549 & 1229 &  &  \\ 
\hline
\end{tabular}
\end{center}
\caption{Number of free coefficients in the ansatz for the set of all $\ell$-loop correlator integrands \p{integrand2} after imposing planarity and discrete symmetry constraints, weak \p{LCOPE} and strong \p{strong} light-cone OPE constraints for different values of the saturation bound $\kappa$ in \p{bound}. We assume the correlator $\vev{{\cal O}_2 {\cal O}_2 {\cal O}_2 {\cal O}_2}$ is already known. In the last column we show the number of additional constraints coming from exponentiation property of the Euclidean OPE for the correlator $\vev{{\cal O}_2 {\cal O}_2 {\cal O}_3 {\cal O}_3}$ in the channel $(14)\to (23)$; they are independent from the light-cone OPE constraints.} \label{tab_ws}
\end{table}

\section{Constraints on Integrated Correlators }\label{condInt}

Using the light-cone OPE relations from the previous section we have greatly simplified the integrands of correlation functions at four and five loops. Meanwhile the integrated four-point functions are given as combinations of four-point conformal integrals. By taking into account their symmetries and relations through magic identities \cite{Drummond:2006rz}, we can see that there is a smaller number of degrees of freedom. For example, while the weak ansatz for the five-loop integrand has 1217 unknown coefficients  at bound $\kappa=5$, the five-loop correlators are labeled by 791 independent coefficients, which we now want to determine using input from integrability.

Henceforth, we will be considering the euclidean OPE limit of the four-point functions, where $u \to 0$ and $v \to 1$. We will assume for simplicity that the lengths of the external operators are such that $L_1\leq L_2$, $L_3 \leq L_4$ and $L_2-L_1\geq L_4-L_3$, since all other cases can be obtained easily with a transformation of the cross-ratios. The OPE decomposition of this correlator is \cite{Nirschl:2004pa}
\begin{align}\label{OPEdecomp}
\langle \mathcal O_{L_1} \mathcal O_{L_2} \mathcal O_{L_3} \mathcal O_{L_4}\rangle =&  \frac{(x_{23}^2)^{L_1-E}(x_{34}^2)^{E-L_3}}{(x_{13}^2)^{L_1}(x_{24}^2)^{L_1+L_2-E}}\frac{(y_{12}^2)^{E} (y_{34}^2)^{L_3}}{(y_{14}^2)^{E-L_1} (y_{24}^2)^{E-L_2}}\times\nonumber\\ 
&\times\sum_{\mathcal O\, \mathrm{with}\, \Delta,S,n,m} C_{12 \mathcal O} \, C_{34 \mathcal O} \,\mathcal G_{\Delta,S}(u,v) \,Y^{(L_1-E,L_2-E)}_{n,m}(\sigma, \tau) \,,
\end{align}
where $2E=L_1+L_2+L_3-L_4$, $Y^{(L_1-E,L_2-E)}_{n,m}$ are the $R$-charge blocks for the $SU(4)$ representation $[n-m,L_4-L_3+2m, n-m]$ and the conformal block takes the following form in the OPE limit \cite{Dolan:2000ut}
\begin{equation}\label{OPEconfblock}
\mathcal G_{\Delta,S}(u,v) \sim u^{\Delta-S} (v-1)^S {}_2 F_1\left(\frac{\Delta+S-L_1+L_2}{2} , \frac{\Delta+S+L_3-L_4}{2};\Delta+S; 1-v\right)  \,.
\end{equation}
The OPE limit is therefore dominated by operators of lowest twist $\Delta-S$ and the $SU(4)$ numbers are restricted such that we have polynomial dependence on the $R$-charge cross-ratios $\sigma$, $\tau$
\begin{align}\label{SU4reps}
n &\in [ E-L_1, \min(E,L_3)]   \,,\nonumber\\
m &\in  [ E-L_1, n]\,.
\end{align}

Meanwhile, from the point of view of the four-point function, we have to sum over a number of R-charge structures, each accompanied by a function of the two spacetime cross-ratios
\begin{equation}\label{CorrDima}
\langle \mathcal O_{L_1} \mathcal O_{L_2} \mathcal O_{L_3} \mathcal O_{L_4} \rangle^{(\ell)} =\lambda^{\ell} C_{L_1 L_2 L_3 L_4} \sum_{\{a_{ij}\}} \tilde F_{\{a_{ij}\}}^{(\ell)}(u,v) \prod_{i<j} (d_{ij})^{a_{ij}} \,,
\end{equation}
where we sum over all $a_{ij}$ such that $\sum_{i\neq j} a_{ij} = L_j$. Not surprisingly, the number of $SU(4)$ representations in \eqref{SU4reps} equals the number of allowed tuples $\{a_{ij}\}$, and one can easily relate them. Notice that there are relations between the functions $\tilde F^{(\ell)}_{\{a_{ij}\}}$ as the correlator must be of the form \eqref{loopCorr}
\begin{equation}\label{Fs}
\tilde F^{(\ell)}_{\{a_{ij}\}} = \sum_{\{b_{ij}\}} \frac{R_{\{a_{ij}-b_{ij}\}} F^{(\ell)}_{\{b_{ij}\}}}{x_{13}^2 x_{24}^2} \,,
\end{equation}
where the non-vanishing $R_{\{\alpha_{ij}\}}$ are the components of $R(1,2,3,4)$ from \eqref{R4} 
\begin{align}
R_{\{ 2,0,0,0,0,2 \}}& =   x_{12}^2 x_{34}^2  \,,&  R_{\{ 1,0,1,1,0,1 \}}&=  x_{13}^2 x_{24}^2 -x_{12}^2 x_{34}^2 - x_{14}^2 x_{23}^2  \,,\nonumber\\
R_{\{ 0,2,0,0,2,0 \} } &=   x_{13}^2 x_{24}^2 \,,& R_{\{ 1,1,0,0,1,1 \}}&= x_{14}^2 x_{23}^2 -x_{12}^2 x_{34}^2 - x_{13}^2 x_{24}^2\,, \nonumber \\
R_{\{ 0,0,2,2,0,0 \}} &=   x_{14}^2 x_{23}^2  \,,& R_{\{ 0,1,1,1,1,0 \} }&=  x_{12}^2 x_{34}^2 -x_{14}^2 x_{23}^2 - x_{13}^2 x_{24}^2 \,.
\end{align}
Each conformally invariant function $F_{\{b_{ij}\}}^{(\ell)}$  is given by a linear combination of conformal integrals (see eq. \p{confint}), which are evaluated in the OPE limit with the method of asymptotic expansions, and they are given as
\begin{equation}\label{confintexp}
F_{\{b_{ij}\}}^{(\ell)} \sim \sum_{k=0}^{\ell} \alpha_k \log^k(u)  \,.
\end{equation}
The unknown coefficients of the integrand enter the functions $F_{\{b_{ij}\}}^{(\ell)} $ as in \eqref{confint}, and each conformal integral can in principle contribute to all powers of $\log^k(u)$, which means that all $\alpha_k$ in \eqref{confintexp} will in principle depend on those unknown coefficients. 
If we look back at the OPE limit of the conformal blocks \eqref{OPEconfblock}, we see that the coefficients multiplying the higher powers of $\log(u)$ contain only lower-loop OPE data. This simple observation has non-trivial consequences, as it implies that those terms can be constrained without difficulty by computing the required lower-loop OPE data with integrability.

\subsection{Constraints from Integrability}\label{intcond}

In order to put constraints on the functions $F_{\{b_{ij}\}}^{(\ell)}$ which enter \eqref{CorrDima}, we must understand what we can say about the equivalent picture of conformal block decomposition. 
Thanks to integrability, we know a lot about the structure of the spectrum \cite{Gromov:2013pga,Gromov:2014caa} and structure constants that enter \eqref{OPEdecomp}. For both quantities the prescriptions are especially tailored for decompactification limits. If an operator has large spin-chain length $L$, then its anomalous dimension is computed with the asymptotic Bethe ansatz. However, when we make $L$ small  the prescription needs to be corrected with finite-size effects, which are given by Luscher corrections. 

\begin{figure}[t]
	\begin{center}
		\includegraphics[scale=.4]{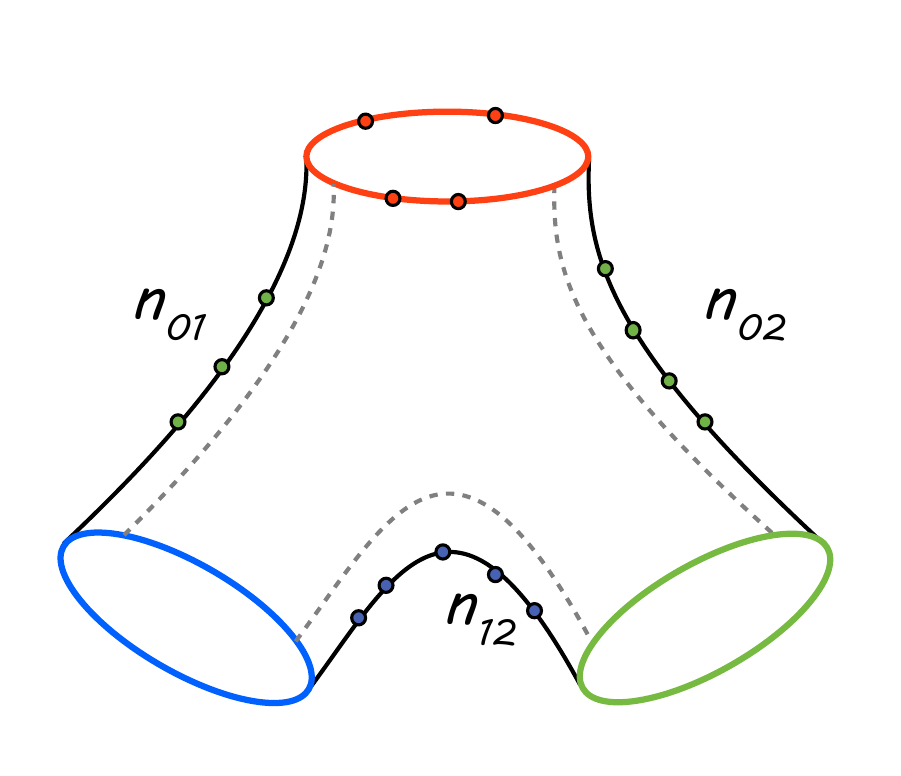}
	\end{center}
	\caption{\normalsize The asymptotic three point function should be suplemented with finite size corrections from the three mirror edges. Following the procedure from \cite{BKV} one is instructed to insert resolution of the identity in each of the edges. The states can have any number of particles on them however the higher the particle number the more surpressed the contribution is. } \label{hexagon} 
\end{figure}

Meanwhile, the OPE coefficients can be computed with Hexagon form factors \cite{BKV}. This method follows a similar expansion, where the decompactification limit is achieved by cutting the pair of pants. This regime is controlled by three parameters, the numbers of tree level Wick contractions between each pair of operators
\begin{equation}\label{lij}
l_{ij}= \frac{1}{2}(L_i + L_j-L_k) \,.
\end{equation}
The asymptotic piece is valid when all $l_{ij}$ are large, but as we decrease the bridge lengths, it must be complemented  with hexagon form factors dressed by $n_{ij}$ virtual excitations in the bridge of length $l_{ij}$, as depicted in Figure \ref{hexagon}. 
For simplicity, let us consider the structure constant between the external operators of length $L_1$ and $L_2$ and an unprotected operator of length $L_0$ that appears in their OPE. It was shown in \cite{3loops} that the contribution of $n_{12}$ virtual excitations in the bottom bridge $l_{12}$ (opposite to the unprotected operator) is suppressed by a factor of
\begin{equation}
g^{2 (n_{12} l_{12}  + n_{12}^2)} \,.
\end{equation} 
This means that even if we put a single virtual excitation in a bridge of length $l_{12}$, the wrapping correction appears at best at $l_{12}+1$ loops. 

We can now use this knowledge when we evaluate the correlator $\langle \mathcal O_{L_1} \mathcal O_{L_2} \mathcal O_{L_3} \mathcal O_{L_4} \rangle^{(\ell)}$. If we pick the contribution of operators $\mathcal O_I$ with $SU(4)$ charges $[M,L_0-2M,M]$ and spins $[S,S]$, at leading twist $\Delta-S=L_0$  the structure constant of those unprotected operators with $\mathcal O_{L_1}$ and $\mathcal O_{L_2}$ is described by hexagons with an opposed bridge of length $l_{12}=1/2( L_1 +L_2 -L_0)$. If we increase the lengths of the external operators to $L_1+n$ and $L_2+n$ and pick again the contribution of the operators $\mathcal O_I$, then we know the structure constants must agree up to $l_{12}$ loops. Or in other words, the OPE limit of the correlators $\langle \mathcal O_{L_1} \mathcal O_{L_2} \mathcal O_{L_3} \mathcal O_{L_4} \rangle^{(\ell)}$ and $\langle \mathcal O_{L_1+n} \mathcal O_{L_2+n} \mathcal O_{L_3} \mathcal O_{L_4} \rangle^{(\ell)}$ must agree for all powers of $\log^k(u)$  with $k \geq \ell-l_{12}$.

We can implement these conditions individually for all different representations in the OPE decomposition of the four-point functions, or equivalently, we can impose them individually on the euclidean OPE limit of the functions $\tilde F_{\{a_{ij}\}}^{(\ell)}$ from \eqref{Fs}. At the end of the day we have
\begin{align}\label{eqs1a}
\left. \left( \tilde F^{(\ell)}_{ \{ n, a_{13}, a_{14}, a_{23}, a_{24}, m \} } -   \tilde F^{(\ell)}_{\{ \ell, a_{13} , a_{14} , a_{23}, a_{24} ,m\} } \right) \right|_{\log^{k\geq \ell-n}} &=0  \quad \mbox{for }n,m>0 \,,\\
\left.\left(\tilde  F^{(\ell)}_{ \{ n, a_{13}, a_{14}, a_{23}, a_{24}, m \} } -  \tilde F^{(\ell)}_{ \{\ell ,a_{13}, a_{14}, a_{23} ,a_{24}, m \}} \right) \right|_{\log^{\ell}} &=0  \quad \mbox{for }\min(n,m)=0 \,.\label{eqs1b}
\end{align}
The reason we treat the case $\min(n,m)=0$ separately is because it corresponds to OPE channels with extremal three-point functions, where there is mixing with double-trace operators. In that case it is not known how to evaluate the OPE coefficients using the integrability methods, so we restrict the constraint to an obvious tree-level statement. 

There is still another set of equations we can impose on the $\tilde F_{\{a_{ij}\}}$, which relates to the fact that opposed wrapping corrections factorize. Apart from a normalization factor $\mathcal N$, the computation of the structure constant requires the evaluation of hexagon form factors  $\mathcal A^{\{l_{ij}\}}_{(n_{01}, n_{12}, n_{02})}$ for different numbers $n_{ij}$ of virtual excitations, where the superscript denotes explicit dependence on some of the bridge lengths $\{l_{ij}\}$. It turns out that the contribution of wrapping on the bottom bridge is always of the form
\begin{equation}
\mathcal A^{(l_{02},l_{12})}_{(n_{01}, n_{12}, n_{02})} = \mathcal A^{(l_{02})}_{(n_{01},0,n_{02})} \mathcal B^{(l_{12})}_{n_{12}} \,,
\end{equation}
which means that the expansion over wrapping corrections factorizes in the following way
\begin{align}\label{factorization}
\mathcal A &= \mathcal A^{(l_{02})}_{(0,0,0)} + \mathcal A^{(l_{02},l_{12})}_{(0,1,0)}+ \mathcal A^{(l_{02})}_{(1,0,0)}+ \mathcal A^{(l_{02})}_{(0,0,1)}+ \mathcal A^{(l_{02})}_{(1,0,1)}+ \mathcal A^{(l_{02},l_{12})}_{(1,1,0)}+ \mathcal A^{(l_{02},l_{12})}_{(0,1,1)}+ \ldots  \nonumber\\
&=  \left( \mathcal A^{(l_{02})}_{(0,0,0)} + \mathcal A^{(l_{02})}_{(1,0,0)}+ \mathcal A^{(l_{02})}_{(0,0,1)} + \mathcal A^{(l_{02})}_{(1,0,1)} + \ldots \right) \left(1+ \mathcal B^{(l_{12})}_{1} + \ldots\right) \nonumber\\
&= \mathcal A^{(l_{02})}\mathcal B^{(l_{12})} \,.
\end{align}

This has important implications for the four-point functions. For example, if the lengths of the external operators are such that $L_2-L_1>L_4-L_3$, then the splittings $l_{02}$ and $l_{04}$ in the structure constants $C_{12 \mathcal O}$ and $C_{34\mathcal O}$ must be distinct. However, thanks to the factorization property \eqref{factorization}, we have\footnote{A naive power counting would imply that $\mathcal A_{(1,1,1)}$ shows up at six loops, but we will prove later that the contribution must be present already at five loops. This must happen through the regularization prescription that is introduced to fix the divergences in $\mathcal A_{(1,0,1)}$, which could in principle invalidate the factorization property. However, at five loops this affects only operators with symmetric splitting, in which case \eqref{factOPE} is trivially satisfied.}
\begin{equation}\label{factOPE}
C^{(l_{02}, l_{12})}_{12 \mathcal O} C^{(l_{04}, l_{34})}_{34 \mathcal O} =  \mathcal N^2 \mathcal A^{(l_{02})}\mathcal B^{(l_{12})} \mathcal A^{(l_{04})}\mathcal B^{(l_{34})}= \mathcal N^2 \mathcal A^{(l_{02})} \mathcal B^{(l_{34})}\mathcal A^{(l_{04})} \mathcal B^{(l_{12})} = C^{(l_{02}, l_{34})}_{1'2' \mathcal O} C^{(l_{04}, l_{12})}_{3'4' \mathcal O} \,,
\end{equation}
where we define new external operators with lengths
\begin{align}
L'_1&= l_{01} + l_{34} \,,& L'_2 &= l_{02}+ l_{34}\,,\nonumber\\
L'_3&= l_{03} + l_{12} \,,& L'_4 &= l_{04}+ l_{12}\,.
\end{align}
Using this insight in the OPE decomposition we can see that the functions $\tilde F_{\{a_{ij}\}}$ must obey the following property
\begin{align}\label{eqs2a}
\tilde F^{(\ell)}_{\{ n, a_{13}, a_{14}, a_{23}, a_{24}, m\} } - \tilde F^{(\ell)}_{\{ m, a_{13} ,a_{14}, a_{23}, a_{24}, n \} }&=0 \quad \mbox{for }n,m>0 \,,\\
\left.\left(\tilde F^{(\ell)}_{ \{ n , a_{13} ,a_{14}, a_{23}, a_{24}, m\} } - \tilde F^{(\ell)}_{ \{ m, a_{13}, a_{14}, a_{23}, a_{24}, n\} }\right)\right|_{\log^{\ell}}&=0 \quad \mbox{for }\min(n,m)=0 \,.\label{eqs2b} 
\end{align}
For the non-extremal case when both $n$ and $m$ are strictly positive, we can impose the equality for all powers of $\log(u)$. Meanwhile, for extremal configurations \eqref{factorization} might not be valid so we restrict the equation to a tree-level statement.\footnote{Interestingly enough, once we fix all four-point functions we observe that both \eqref{eqs1b} and \eqref{eqs2b} would be valid if applied to the same $\log(u)$ powers of \eqref{eqs1a} and \eqref{eqs2a}.}

Let us remark that even though we used knowledge from integrability to formulate equations \eqref{eqs1a}, \eqref{eqs1b}, \eqref{eqs2a} and \eqref{eqs2b}, they require absolutely no numerical input from integrable machinery, and yet they introduce  powerful constraints on the four-point functions.

\subsection{OPE data in the $\mathfrak{sl}(2)$ sector}
\label{sl2}

In the previous subsection we derived constraints on the functions $ F_{\{b_{ij}\}}^{(\ell)}$ by looking at the integrability description of three-point functions and using the knowledge of when opposed wrapping corrections first start to kick in. This nice exercise allows us to fix many of the unknown coefficients without having to do any actual computation with the integrability machinery.
In this section we explain how to further constrain the integrand by computing the simplest components of three-point functions in the $\mathfrak{sl}(2)$ sector.

By choosing specific polarization vectors $y_i$ for the external protected operators, we can single out the OPE channel in \eqref{OPEdecomp} with $SU(4)$ charges $[0,L,0]$, twist $L$ and spin $S$. These are operators of the form
\begin{equation}
\mathrm{Tr}[Z D^S Z^{L-1}] + \ldots 
\end{equation}
and correspond to spin-chain excitations in the $\mathfrak{sl}(2)$ sector. This is an especially easy sector within the integrability framework, where we can find all solutions to the Bethe equations without difficulty. Since this is a rank-one sector, it is also a relatively easy setup for the computation of structure constants.

In order to pick such  an OPE channel we should analyze correlators of the form 
\begin{equation}\label{CorrSL2}
\langle \mathrm{Tr}[X^{l_{12}} Z^{l_{01}}](x_1) \mathrm{Tr}[\bar X^{l_{12}} Z^{l_{02}}](x_2) \mathrm{Tr}[ Y^{l_{34}} \bar Z^{l_{03}}](x_3) \mathrm{Tr}[\bar  Y^{l_{34}} \bar Z^{l_{04}}](x_4)\rangle \,,
\end{equation}
at the leading power of $u^{-l_{12}}$. In terms of the polarization vectors this can be achieved by choosing \cite{Vieira:2013wya}
\begin{align} \label{polarizations}
y_1&= \frac{1}{\sqrt 2}(1, i ,\alpha_1, i \alpha_1,0,0)  \,,& y_2&=\frac{1}{\sqrt 2}(1, i ,\alpha_2,- i \alpha_2,0,0)  \,,\nonumber\\
y_3&= \frac{1}{\sqrt 2}(1, -i ,0,0,\alpha_3, i \alpha_3)  \,,& y_4&=\frac{1}{\sqrt 2}(1, -i ,0,0,\alpha_4,- i \alpha_4)  \,,
\end{align}
and then taking derivatives of the correlator 
\begin{equation}\label{correlators}
\left.\frac{1}{(l_{12}! l_{34}!)^2}\left(\frac{\partial}{\partial \alpha_1} \frac{\partial}{\partial \alpha_2}\right)^{l_{12}} \left(\frac{\partial}{\partial \alpha_3} \frac{\partial}{\partial \alpha_4}\right)^{l_{34}}\langle \mathcal O_{L_1}(y_1) \mathcal O_{L_2}(y_2)\mathcal O_{L_3}(y_3) \mathcal O_{L_4}(y_4)\rangle \right|_{\alpha_i=0}\,.
\end{equation}

In terms of the four-point function \eqref{CorrDima}, we are picking the contribution of a subset of the functions $\tilde F_{\{a_{ij}\}}^{(l)}$ which are of the form
\begin{equation}\label{FsSL2}
\sum_{\alpha=0}^{l_{01}}\frac{\tilde F^{(\ell)}_{\{l_{12},\alpha, l_{01}-\alpha, l_{03}-\alpha,l_{02}-l_{03}+\alpha,l_{34}\}}}{u^{l_{12}} v^{l_{03}-\alpha}} = \sum_{\beta=0}^{l_{01}-1} \frac{(1-v)^2 F^{(\ell)}_{\{l_{12}-1,\beta,l_{01}-1-\beta,l_{03}-1-\beta,l_{02}-l_{03}+\beta,l_{34}-1\}}}{u^{l_{12}} v^{l_{03}-\beta}} \,.
\end{equation}
Notice that only two elements of $R$ contribute for the right-hand side of \eqref{FsSL2}, namely $R_{\{1,1,0,0,1,1\}}$ and $R_{\{1,0,1,1,0,1\}}$. This happens because $R_{\{2,0,0,0,0,2\}}$ is always subleading in $u$, while the other three terms $R_{\{0,\ldots,0\}}$ happen to be subleading for the specific polarizations chosen.

In this way we are able to extract sum rules for operators in the $\mathfrak{sl}(2)$ sector, which we now want to match with sum rules obtained from integrability. By equating them we will be able to determine many of the unknown coefficients in the functions $F^{(\ell)}_{\{b_{ij}\}}$. 

The required three-point functions are obtained by a finite-volume correlator of two hexagon operators. This is a hard object to obtain and so one considers the two-point function of the hexagon operators as an expansion around the infinite-volume limit. This is particularly useful at a perturbative level where the finite-volume effects can be tamed order by order in the coupling. 
\begin{figure}[t]
	\begin{center}
		\includegraphics[scale=.32]{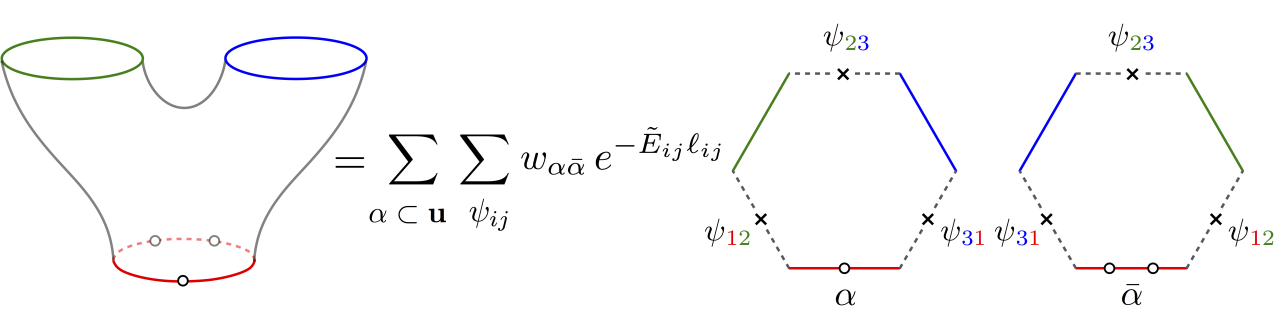}
	\end{center}
	\caption{As we cut the pair of pants in two hexagons, we must partition the Bethe roots $\mathbf u$ into the sets $\alpha$ and $\bar \alpha$ which populate the physical edge of each of the hexagon form factors. Finite-size corrections are obtained by inserting particle/anti-particle pairs in the mirror edges of the hexagons, denoted here by $\psi_{ij}$.} \label{Hexagons} 
\end{figure}
Each non-protected operator is represented by its Bethe roots, which are distributed among the two hexagons\footnote{Notice that one should sum over all possible ways of distributing the Bethe roots among the two hexagons.}.  The infinite-volume expansion corresponds to inserting a resolution of the identity in each unphysical edge of the hexagon, which in practice is written as an infinite sum of virtual excitations (including the term with zero particles). A schematic representation of this proposal is portrayed in Figure \ref{Hexagons}. The creation and propagation of the virtual excitations costs energy, so their contribution appears at higher orders in perturbation theory. The explicit coupling dependence of different finite-size corrections can be found in \cite{3loops}.

We will consider a ratio of structure constants, where the numerator is the OPE coefficient for a non-protected operator of length $L_0$ in the $\mathfrak{sl}(2)$ sector with two protected operators of lengths $L_1$ and $L_2$, while the denominator corresponds to the structure constant for three protected operators of lengths $L_0$, $L_1$ and $L_2$ 
\begin{equation}\label{c123}
\frac{C^{\bullet\circ\circ}}{C^{\circ\circ\circ}}= \sqrt{\frac{\prod_k\mu(u_k)}{\langle \{u\} | \{u\}\rangle \prod_{i<j} S(u_i,u_j)}} \; \mathcal A \,,
\end{equation}
where $\langle \{u\} | \{u\}\rangle $ is the Gaudin norm, $\mu$ is the measure which controls the asymptotic normalization of one-particle states,  $S$ is the $\mathfrak{sl}(2)$ S-matrix and $\mathcal A$ is the two-point function of hexagon operators.
In this work it was sufficient to consider the asymptotic hexagon form factors $\mathcal A_{(0,0,0)}$ and the single-particle wrapping correction in the opposed mirror channel $\mathcal A_{(0,1,0)}$, which we now review.

\subsubsection*{Asymptotic contribution}
The leading asymptotic contribution to the hexagon form factors is \cite{BKV}
\begin{equation} \label{A000}
\mathcal A_{(0,0,0)}=\sum_{\alpha \cup \bar \alpha = \{u\}} (-1)^{|\bar{\alpha}|}\omega(\alpha, \bar{\alpha}) h(\alpha) h(\bar \alpha)\,,
\end{equation}
where $\omega(\alpha, \bar{\alpha})$ is the splitting factor
\begin{equation}
\omega(\alpha , \bar \alpha)=\prod_{i \in \bar{\alpha}}e^{i p(u_i) l_{02} } \prod_{j\in \alpha, i>j} S(u_i,u_j)
\end{equation}
and $h(u)$ the hexagon form factor for a set of excitations $\{u\}$ in a single physical edge
\begin{equation}
h(u)= \prod_{i<j} \frac{x^-(u_i) - x^-(u_j)}{x^- (u_i) - x^+(u_j)} \frac{1-\frac{1}{x^-(u_i) x^+(u_j)}}{1-\frac{1}{x^+(u_i) x^+(u_j)}} \frac{1}{\sigma(u_i,u_j)} \,,
\end{equation}
where $x^\pm$ are the Zhukowsky variables and $\sigma$ is the BES dressing phase.

\subsubsection*{Finite-size corrections}\label{Opposing}
The computation of the hexagon with a single virtual excitation in the mirror edge opposed to the unprotected operator boils down to the evaluation of the following integral \cite{BKV}
\begin{equation}\label{opposed}
\mathcal{A}_{(0, 1, 0)} = \mathcal{A}_{(0,0,0)}\sum_{a\geq 1}\int \frac{du}{2\pi} \mu_{a}(u^{\gamma})
e^{ip_{a}(u^{\gamma})l_{12}} \frac{T_{a}(u^{-\gamma})}{h_{1a}(\textbf{u}, u^{-\gamma})}\,,
\end{equation}
where $l_{12}$ is the length of the opposed bridge, $T_{a}$ is the transfer matrix, $h_{1a}$ the hexagon form factor and $\mu_a(u^{\gamma})$ the mirror measure for a bound state of $a$ derivatives, see \cite{Basso:2017muf} for the precise definition of each of these factors. It is instructive to show the leading order expansion of the integral at weak coupling
\begin{align}
\frac{a}{(u^2+\frac{a^2}{4})^{2+l_{23}}}\frac{Q(u^{[a+1]})+Q(u^{[-a-1]})-Q(u^{[a-1]})-Q(u^{[-a+1]})}{Q(i/2)}\, ,
\end{align}
where $Q(u)=\prod_{i}(u-u_i)$ is a polynomial of degree $M$ and $u_i$ are the $M$ Bethe roots for the state under consideration. Notice that the integral in $u$ is divergent for small $l_{12}$ and large enough $M$. As explained in \cite{3loops}, the sum over bound states $a$ cures this divergence, but it is technically hard to perform the sum before the integration in $u$. It was then shown that \eqref{opposed} can be evaluated efficiently with the following method:
\begin{itemize}
	\item Consider the function $Q(u)=e^{iut}$;
	\item Do the integral in $u$ by residues;
	\item Write the result of the integration  in terms of nested harmonic sums;
	\item Perform the remaining sums by identifying it with harmonic polylogarithms.
\end{itemize}
The original polynomial can be recovered by acting with $Q(-i\partial_t)$ in the final result. The advantage of using the plane-wave $e^{iut}$ is that it makes the integral more convergent, allowing the evaluation of the integral in $u$ by residues. The sum over bound states is trivialized once one identifies the sum as harmonic polylogarithms. Another advantage is that this method gives at once the finite-size contribution for any state.

\subsection{Consistency conditions}
\label{secwrappings}

While the data from asymptotic hexagons and opposed wrapping can introduce strong constraints on the undetermined coefficients, there are certainly many configurations in the $\mathfrak{sl}(2)$ sector which also require the evaluation of adjacent wrappings. It is however possible to fix coefficients that appear in such configurations without evaluating any adjacent wrapping explicitly, and we will also see how the input of the opposed wrapping correction to $(\ell-2)$ loops will help constrain the $\ell$-loop four-point functions.

\begin{figure}[t]
	\begin{center}
		\includegraphics[scale=.2]{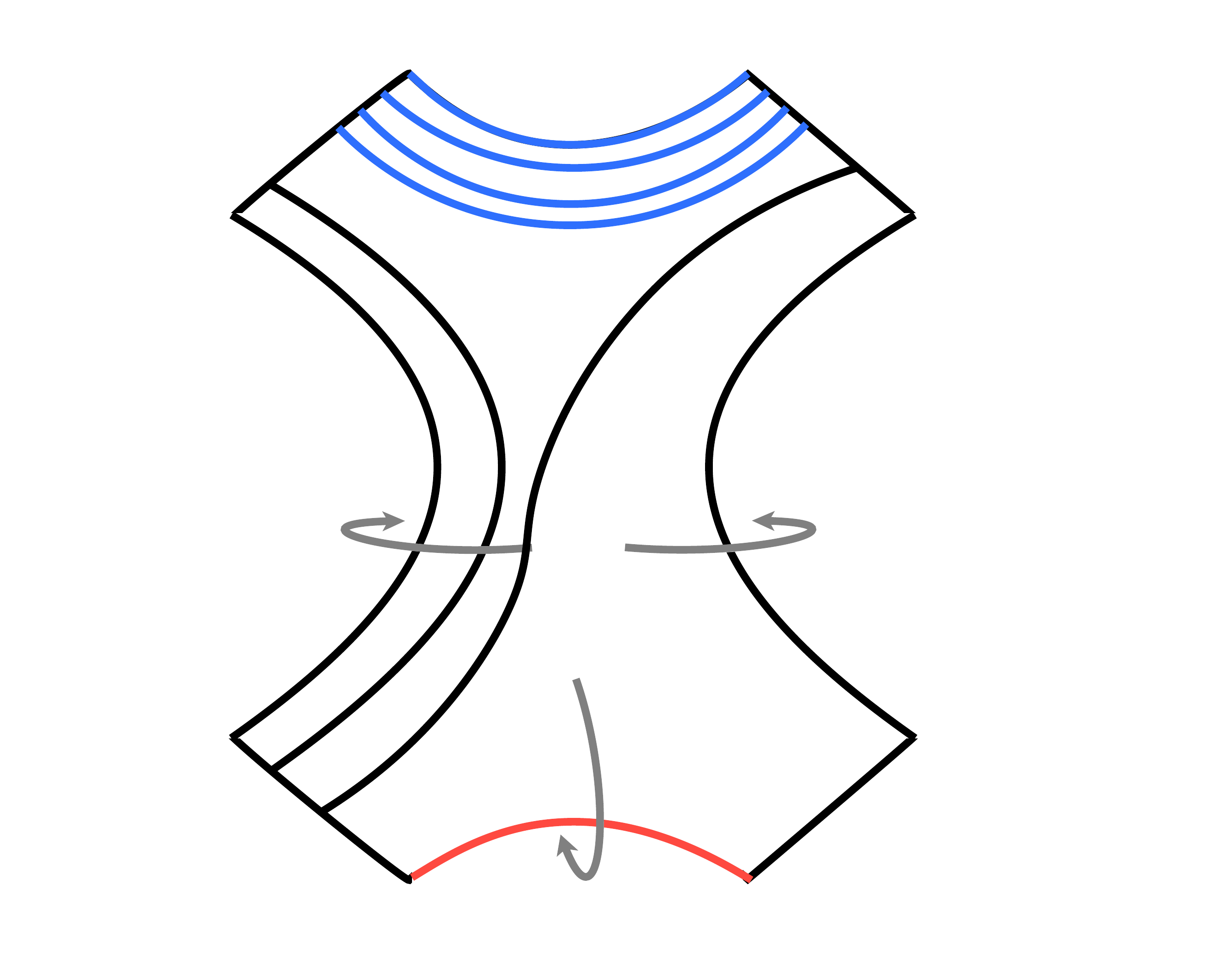}
	\end{center}
	\caption{\normalsize Example of a four-point correlator where blue lines denote contraction between $X$ and $\bar X$, red lines connect $Y$ and $\bar Y$ and all other lines correspond to $Z$ and $\bar Z$ fields. The leading operators in the OPE limit have twist 4, and we get a product of distinct structure constants. The OPE coefficient in the top has equal splitting between adjacent bridges and opposed bridge of length 4, so that virtual excitations in those lines do not contribute at four loops. Meanwhile the OPE coefficient in the bottom half has an assymetric split between left and right adjacent bridges, and the opposed bridge has length one, which implies the appearance of opposed wrapping at two loops.} \label{hexagon4pt} 
\end{figure}

Once we take the OPE limit of the correlators it is simple to extract sum rules $ P^{(\ell,n)}$ which are defined by
\begin{equation}
\sum_{\mathcal O } C_{12 \mathcal O}(\lambda) C_{34 \mathcal O}(\lambda) e^{\gamma_{\mathcal O}(\lambda) \,z}= \sum_{\ell=0}^\infty \sum_{a=0}^\ell \lambda^\ell z^a  P^{(\ell,a)} \,,
\end{equation}
where $\gamma_{\mathcal O}$ are the anomalous dimensions and we sum over all operators $\mathcal O$ with given dimension, spin and $SU(4)$ charges.
For simplicity, let us now focus on a four-loop example. If we look at correlators with different weights then we can extract sum rules for configurations where the unprotected operator has different splittings $l_{01}$ and $l_{03}$ and the opposed bridge lengths have values $l_{12}$ and $l_{34}$
\begin{align}
P_{(l_{12},l_{34},l_{01},l_{03})}^{(4,0)}= \sum_{I}& C_{l_{12},l_{01},I}^{(0)} C_{l_{34},l_{03},I}^{(4)} + C_{l_{12},l_{01},I}^{(1)} C_{l_{34},l_{03},I}^{(3)}  + C_{l_{12},l_{01},I}^{(2)} C_{l_{34},l_{03},I}^{(2)} \nonumber\\
&+C_{l_{12},l_{01},I}^{(3)} C_{l_{34},l_{03},I}^{(1)}  + C_{l_{12},l_{01},I}^{(4)} C_{l_{34},l_{03},I}^{(0)}  \,,
\end{align}
with $C_{l_{ij},l_{0k},I}^{(\ell)}$ the $\ell$-loop OPE coefficient for opposed bridge of length $l_{ij}$, adjacent bridge length $l_{0k}$ and operator $\mathcal O_I$ with the correct dimension, spin and $SU(4)$ charges. 
This type of sum rule can be extracted from the analysis of correlators like the one depicted in Figure \ref{hexagon4pt}.

As explained above, the opposed wrapping contributions factorize in the computation of the structure constant, so we can rewrite it as
\begin{equation}
C_{l_{ij},l_{0k},I}^{(\ell)}= \mathcal N_I \left(\mathcal A_{\mathrm{adj},I}^{(\ell,l_{0k})} +  \sum_{k=2}^\ell\mathcal A_{\mathrm{adj},I}^{(\ell-k,l_{0k})} \mathcal B_{1,I}^{(k,l_{ij})}\right) \,,
\end{equation}
where $\mathcal N_I$ denotes the normalization factor from \eqref{c123}, $\mathcal A_{\mathrm{adj},I}^{(\ell,l_{0k})}$ is the sum of the asymptotic and adjacent wrapping contributions at $\ell$ loops for adjacent bridge of length $l_{0k}$, and $\mathcal B_{1,I}^{(\ell,l_{ij})}$ is the $\ell$-loop single-particle opposed wrapping for opposed bridge length $l_{ij}$. 
For the configuration when both opposed bridges have length four, in which case there are only adjacent wrappings, we have
\begin{align}\label{consistency0}
P_{(4,4,l_{01},l_{03})}^{(4,0)} = \sum_{I}\mathcal N_I^2  &\left(\mathcal A_{\mathrm{adj},I}^{(0,l_{01})} \mathcal A_{\mathrm{adj},I}^{(4,l_{03})} + \mathcal A_{\mathrm{adj,I}}^{(1,l_{01})} \mathcal A_{\mathrm{adj},I}^{(3,l_{03})}+ \mathcal A_{\mathrm{adj},I}^{(2,l_{01})} \mathcal A_{\mathrm{adj},I}^{(2,l_{03})}\right. \nonumber\\
&\quad\left.+ \mathcal A_{\mathrm{adj},I}^{(3,l_{01})} \mathcal A_{\mathrm{adj},I}^{(1,l_{03})}+\mathcal A_{\mathrm{adj},I}^{(4,l_{01})} \mathcal A_{\mathrm{adj},I}^{(0,l_{03})} \right)\,.
\end{align}
As we lower the length of the opposed bridge to $l_{34} < 4$, we must add contributions from opposed wrapping, which starts at two loops, so we have
\begin{align}\label{consistency1}
& P_{(4,l_{34},l_{01},l_{03})}^{(4,0)} = P_{(4,4,l_{01},l_{03})}^{(4,0)}+\sum_{I} \mathcal N_I^2 \left(\mathcal A_{\mathrm{adj},I}^{(2,l_{01})} \mathcal A_{\mathrm{adj},I}^{(0,l_{03})} +\mathcal A_{\mathrm{adj},I}^{(1,l_{01})} \mathcal A_{\mathrm{adj},I}^{(1,l_{03})} +\mathcal A_{\mathrm{adj},I}^{(0,l_{01})} \mathcal A_{\mathrm{adj},I}^{(2,l_{03})} \right) \mathcal B_{1,I}^{(2,l_{34})} \nonumber\\
&+\sum_I \mathcal N_I^2 \left(\mathcal A_{\mathrm{adj},I}^{(1,l_{01})} \mathcal A_{\mathrm{adj},I}^{(0,l_{03})} +\mathcal A_{\mathrm{adj},I}^{(0,l_{01})} \mathcal A_{\mathrm{adj},I}^{(1,l_{03})} \right) \mathcal B_{1,I}^{(3,l_{34})} +\sum_I \mathcal N_I^2 \mathcal A_{\mathrm{adj},I}^{(0,l_{01})} \mathcal A_{\mathrm{adj},I}^{(0,l_{03})} \mathcal B_{1,I}^{(4,l_{34})}.
\end{align}
Notice that the adjacent wrapping corrections can only start at three loops, which means that $\mathcal A_{\mathrm{adj},I}$ always simplifies to the asymptotic contribution in \eqref{consistency1}. Therefore the only unknowns are the opposed wrappings $\mathcal B_{1,I}^{(\ell,l_{34})}$, but we obtain an overconstrained system of equations because they appear in sum rules for different splittings $l_{01}$ and $l_{03}$.  In the $\mathfrak{sl}(2)$ sector there are $\lfloor L/2 \rfloor$ operators of twist $L$ and spin 2, while there are $1/2(\lfloor L/2\rfloor +  \lfloor L/2\rfloor^2 )$ configurations for the splitting of the twist $L$ operator in the four-point function. This poses non-trivial constraints on the undetermined coefficients of the four-point correlators.

Furthermore, if we let both opposed bridges become smaller, with $l_{12},l_{34} < 4$, then the sum rule is 
\begin{align}\label{consistency2}
P_{(l_{12},l_{34},l_{01},l_{03})}^{(4,0)} =& P_{(4,l_{34},l_{01},l_{03})}^{(4,0)}  + P_{(l_{12},4,l_{01},l_{03})}^{(4,0)} - P_{(4,4,l_{01},l_{03})}^{(4,0)}\nonumber\\
&+ \sum_{I} \mathcal N_I^2\mathcal A_{(0,0,0),I}^{(0,l_{01})} \mathcal A_{(0,0,0),I}^{(0,l_{03})} \mathcal B_{1,I}^{(2,l_{12})} \mathcal B_{1,I}^{(2,l_{34})} \,.
\end{align}
We can see that it is related to the sum rules in \eqref{consistency0} and \eqref{consistency1}, and these relations can be easily implemented with the knowledge of relatively simple objects: asymptotic hexagon form factors and opposed wrapping at two loops. Moreover, if any of the opposed bridges has length bigger than one, then the last term in \eqref{consistency2} is identically zero. The fact that  sum rules for different opposed bridge lengths respect such relations imposes non-trivial constraints on the four-point correlators. Finally, at higher loops the arguments are very similar, with the only difference being that at $\ell$ loops the last term in \eqref{consistency2} will include opposed wrapping corrections up to $(\ell-2)$ loops and $\mathcal A_{\mathrm{adj},I}$ in \eqref{consistency1} might include the contribution of adjacent wrapping corrections.

\section{Results} \label{res_sec}

In this section we apply the methods described above in order to fix all four- and five-loop four-point functions of protected operators. Since we could not prove the validity of the stronger version of the light-cone OPE relations \eqref{strong} above three loops, we shall always start from the integrand constrained only by the weak relations of \eqref{LCOPE}. 

We need to obtain the functions $F^{(\ell)}_{\{b_{ij}\}}$ for all indices $b_{ij}$ ranging between 0 and $(\ell-1)$. While this bound was proved up to three loops, we do not have a direct proof at higher loops, but its existence is natural from the point of view of Feynman diagrams. At any loop order there is a maximum number of fields that can be involved in a given interaction vertex, which means that for large enough operators there will always be a number of spectator fields. Furthermore, our results seem to indicate that the strong light-cone OPE relations \eqref{strong} are valid at four and five loops, and the strong version of the integrand is the same for all values of the bound larger or equal than $\kappa_\mathrm{min}(\ell)$, which seems to indicate that is the correct bound.

\subsection{Four Loops}

At four loops we expect the bound on the $\{ b_{ij} \}$ in eq.~\p{bound} to be $\kappa=3$, but in order to test this we start with functions $F^{(4)}_{\{b_{ij}\}}$ whose indices are bounded at $\kappa=5$. The weak  ansatz fixes all 2451 functions up to 149 undetermined coefficients, which is also the number of degrees of freedom in the integrated correlators. 

If we impose the equations from section \ref{intcond}, we are able to fix 130 of the 149 coefficients. Then we consider correlators in the $\mathfrak{sl}(2)$ sector by analyzing the configurations from \eqref{CorrSL2}. If the adjacent bridge length is $l_{01}$ and the opposed bridges have lengths $l_{12}$ and $l_{34}$, then the asymptotic hexagons are the only contribution up to $\min(l_{12},l_{34},l_{01}+1)$ loops. That means that we can compare the data obtained with all $\log^k(u)$ terms of the correlator for $k\geq 4-\min(l_{12},l_{34},l_{01}+1)$. There is a remarkable amount of information and we are able to determine 18 coefficients in this way.  At this point the integrand is completely fixed up to a single coefficient, which we determine using the consistency conditions presented in section \ref{secwrappings}. We need to evaluate opposed wrapping up to two loops, and by comparing sum rules for different opposed bridge lengths we are able to fix the last coefficient.

In the end, we are able to fix all planar four-loop four-point functions with striking ease. Regarding the result obtained, it is very interesting to observe that the bound on the indices $\{ b_{ij} \}$ does turn out to reduce to $\kappa = 3$. Moreover, we find that the solution to the weak version of light-cone OPE \eqref{LCOPE} is consistent with the strong criterion \eqref{strong}. We also evaluated all three- and four-loop opposed wrapping corrections for spin 2 operators up to twist 20 and obtained a perfect match with the data extracted from the four-point function.

\subsection{Five Loops}

At five loops we expect the bound on the $\{ b_{ij}\}$ from eq.~\p{bound} to be $\kappa=4$, but once again we test this conjecture by starting with the bound $\kappa=5$. We need to consider 2451 functions $F^{(5)}_{\{b_{ij}\}}$, which contain 1217 undetermined coefficients, but when we consider symmetries of the conformal integral and magic identities between them we can show that the integrated correlator depends only on 791 coefficients.

At five loops it is quite difficult to take the OPE limit of the conformal integrals, so only the order  $(1-v)^0$ of the expansions is available. That means that if we naively take $v$ to one in the conditions of section \ref{intcond} then we might lose some important information. This happens because the $SU(4)$ representations $[M,L_0-2M,M]$ that appear in a given correlator at twist $L_0$ are combinations of the functions
\begin{equation}\label{L5careful}
\tilde F^{(5)}_{\{ l_{12}, \alpha, l_{01}-\alpha, l_{03}-\alpha, l_{02}-l_{03}+\alpha,l_{34} \} } \,,
\end{equation}
for $0\leq\alpha\leq l_{01} $. It is easy to see that the numbers match if one remembers that only representation with $L_0-2M\geq L_2-L_1$ are allowed, or equivalently, $M\leq l_{01}$. Since the representation $[0,L_0,0]$ corresponds to operators in the $\mathfrak{sl}(2)$ sector, we know that the first non-protected operator has spin two and therefore the representation must come with a factor of $(1-v)^2$. Analogously, the representation $[1,L_0-2,1]$ will always come with a factor of $(1-v)$, which means that there are two linear combinations of the functions \eqref{L5careful} that will be vanishing at $v=1$. In order to obtain a maximum number of constraints from \eqref{eqs1a}, \eqref{eqs1b}, \eqref{eqs2a} and \eqref{eqs2b} we must then find what those linear combinations are and substitute the expansions of the conformal integrals at the leading non-vanishing order of those equations. 
Once we take this into consideration, we are able to fix 578 of the 791 undetermined coefficients.

Then, just like at four loops, we can consider the data from asymptotic hexagon form factors and compare with the $\log^k(u)$ terms of the correlator for $k\geq 5-\min(l_{12},l_{34},l_{01}+1)$, which fixes 70 more coefficients. At this point we use the technique introduced in section \ref{secwrappings}, where we extract adjacent wrapping corrections by looking at correlators with opposed bridges of length 5, and then look for consistent conditions on the data of lower opposed bridge lengths. This proves very effective, and we are able to fix a further 120 coefficients by inputing only two- and three-loop opposed wrapping effects.

At this point we have fixed all correlators up to 23 coefficients. In order to fix those last degrees of freedom, we look again at equations \eqref{eqs1a} and \eqref{eqs2a}, but in terms of conformal integrals and not their OPE expansions. For each equation we must consider only the conformal integrals which can contribute at the relevant powers of $\log(u)$, and once we do that we notice that all equations at this point depend only on four distinct conformal integrals
\begin{align}
I_1 &= \int \frac{\mathrm d^4 x_5  \mathrm d^4 x_6  \mathrm d^4 x_7  \mathrm d^4 x_8 \mathrm d^4 x_9 \;\;\;x_{14}^2 x_{23}^4 x_{24}^2}{x_{15}^2 x_{16}^2  x_{25}^2  x_{26}^2  x_{27}^2  x_{28}^2  x_{35}^2  x_{37}^2  x_{39}^2  x_{46}^2  x_{48}^2  x_{49}^2  x_{57}^2  x_{68}^2  x_{79}^2  x_{89}^2 } \,,\nonumber\\
I_2 &= \int \frac{\mathrm d^4 x_5  \mathrm d^4 x_6  \mathrm d^4 x_7  \mathrm d^4 x_8 \mathrm d^4 x_9 \;\;\; x_{13}^2 x_{14}^2 x_{23}^2 x_{24}^2 x_{29}^2}{x_{15}^2 x_{17}^2  x_{19}^2  x_{25}^2  x_{26}^2  x_{27}^2  x_{28}^2  x_{36}^2  x_{38}^2  x_{39}^2  x_{45}^2  x_{46}^2  x_{49}^2  x_{56}^2  x_{78}^2  x_{79}^2 x_{89}^2} \,,\nonumber\\
I_3 &= \int \frac{\mathrm d^4 x_5  \mathrm d^4 x_6  \mathrm d^4 x_7  \mathrm d^4 x_8 \mathrm d^4 x_9 \;\;\; x_{12}^2 x_{13}^2 x_{14}^2 x_{23}^2 x_{24}^2 x_{29}^2 x_{59}^2}{x_{15}^2 x_{16}^2 x_{18}^2  x_{19}^2  x_{25}^2  x_{27}^2  x_{28}^2  x_{29}^2  x_{35}^2  x_{36}^2  x_{37}^2  x_{46}^2  x_{47}^2  x_{49}^2  x_{56}^2  x_{58}^2  x_{79}^2 x_{89}^2}\,, \nonumber\\
I_4 &= \int \frac{\mathrm d^4 x_5  \mathrm d^4 x_6  \mathrm d^4 x_7  \mathrm d^4 x_8 \mathrm d^4 x_9 \;\;\; x_{13}^4 x_{24}^4}{x_{15}^2 x_{16}^2  x_{18}^2  x_{25}^2  x_{26}^2  x_{27}^2  x_{35}^2  x_{37}^2  x_{39}^2  x_{45}^2  x_{48}^2  x_{49}^2  x_{67}^2  x_{68}^2  x_{79}^2  x_{89}^2 } \,. \label{confintex}
\end{align}
Since $I_2$ and $I_4$ are products of one- and four-loop conformal integrals, we can easily obtain their expansions to order $(1-v)^4$. Meanwhile $I_1$ and $I_3$ are genuine five-loop integrals but luckily they are some of the simpler ones and we were able to perform the asymptotic expansions to order $(1-v)^1$. By plugging these new expansions back in the equations we were able to obtain new constraints corresponding to higher spin contributions in the OPE decomposition of the four-point correlators, which fixed all but one of the coefficients.

Finally, we consider the correlator
\begin{equation}
\mathcal G=(x_{13}^2)^{2p} (x_{24}^2)^{2p}\langle \mathrm{Tr}[Z^p \bar Y^p](x_1)  \mathrm{Tr}[Y^p \bar X^p](x_2)  \mathrm{Tr}[\bar Z^{2p}](x_3) \mathrm{Tr}[Z^p X^p](x_4)\rangle \,,
\end{equation}
which we evaluate with equation \eqref{CorrDima}, leading to
\begin{equation}\label{FrankCorr}
\mathcal G^{(5)}= u^{-p}\left( F^{(5)}_{ \{ p,p-2,0,0,p-2,p \} }+(v-1)F^{(5)}_{ \{ p-1,p-1,0,0,p-1,p-1 \}}\right) + \mathcal O(u^{-p+1})\,.
\end{equation}
If $p\geq7$, then both functions on the right-hand side of \eqref{FrankCorr} saturate the bound and we have at leading order in $u$
\begin{equation}
\mathcal G^{(5)}= \frac{v}{u^p} F^{(5)}_{ \{ 5,5,0,0,5,5 \} } \,,
\end{equation}
for which all orders of $\log(u)$ depend on the last undetermined coefficient. Thankfully this correlator has been evaluated in the regime of large $p$ through hexagonalization\footnote{We thank Frank Coronado for sharing this result prior to publication.} \cite{Frank} and we can in this way fix all planar five-loop four-point functions.

It is interesting to note that the solution to the weak ansatz of the integrand is compatible with the strong light-cone OPE relations \eqref{strong} and the bound on the indices $\{ b_{ij} \}$ does reduce to $\kappa=4$ as expected. We also evaluated all four-loop opposed wrapping corrections for spin 2 operators up to twist 20 and once again obtained a perfect match with the data extracted from the five-loop four-point function.

\subsection{Triple Wrapping}

As mentioned above, the integrability approach to the computation of three-point functions depends on an asymptotic contribution and finite-size corrections. By considering specific polarizations and/or large enough external operators, one can postpone some of the wrapping corrections to higher loops and in some cases even isolate specific finite-size corrections. 

A simple example where this happens comes from considering the following family of four-point functions
\begin{align}
\langle \mathcal{O}_2(x_1)\mathcal{O}_2(x_2)\mathcal{O}_n(x_3) \mathcal{O}_n(x_4)\rangle \,,
\end{align}
where $n\ge 2$. Looking at the singlet $SU(4)$ representation in the OPE limit of small $u$ and $(1-v)$  probes the product of structure constants $C_{22\mathcal{K}}C_{nn\mathcal{K}}$ where $\mathcal{K}$ represents the Konishi operator.  As we increase the length $n$ of the operators, the wrapping corrections in the adjacent bridges remain the same, but the contribution of the virtual excitation in the opposed bridge is delayed to $n$ loops. For example, by looking at the configuration where $n$ is six we are able to extract the contribution of adjacent wrappings $\mathcal A_{\mathrm{adj}}= \mathcal A_{(1,0,0)}+\mathcal A_{(0,0,1)}+\mathcal A_{(1,0,1)}$ to the structure constant
\begin{align}
&\mathcal A_{\mathrm{adj}} =\lambda^3 \left(324+864 \zeta_3-1440 \zeta_5\right)-\lambda^4 \left(9801+648 \zeta_3+9360 \zeta_5+3888 \zeta_3^2-27720 \zeta_7\right) \nonumber\\
&\quad+\lambda^5\left(217080-154224 \zeta_3+139536 \zeta_5-10368 \zeta_3^2+91980 \zeta_7+116640 \zeta_3 \zeta_5-435456 \zeta_9\right) \,.
\end{align}
Perhaps more interestingly, we can now evaluate the difference of sum rules introduced in \eqref{consistency1}
\begin{equation}
P_{(1,l_{34},1,1)}^{(5,a)}- P_{(1,5,1,1)}^{(5,a)}
\end{equation}
which probe the one-particle contribution to the bottom edge. For opposed bridge lengths $2 \leq l_{34} \leq 4$ these correlators exactly match the opposed wrapping contributions (we use the notation introduced earlier $\mathcal A^{(l_{34})}_{(0,1,0)}~=\mathcal A_{(0,0,0)}\mathcal B^{(l_{34})}_1$)
\begin{align}
\mathcal B_1^{(2)} &=  \lambda^3 \left(120 \zeta_5 \right)+ \lambda^4 \left(240 \zeta_5-216 \zeta_3^2-2100 \zeta_7 \right) \nonumber\\
&\qquad +\lambda^5 \left(-2400 \zeta_5 +432 \zeta_3^2-3507 \zeta_7+5400 \zeta_5 \zeta_3+31752 \zeta_9 \right)\,, \nonumber\\
\mathcal B_1^{(3)} &=   \lambda^4 \left(420 \zeta_7 \right)  + \lambda^5 \left( 840 \zeta_7-1080 \zeta_3 \zeta_5-10584 \zeta_9\right) \,,\nonumber\\
\mathcal B_1^{(4)} &=  \lambda^5 \left(1512 \zeta_9 \right)\,.
\end{align}
On the other hand, at $l_{34}=1$ there is a mismatch with the wrapping correction
\begin{align}
\mathcal B_1^{(1)}=& \lambda^2 \left( 	36 \zeta_3\right) + \lambda^3 \left(72 \zeta_3-360 \zeta_5 \right) + \lambda^4 \left( -720 \zeta_3+432 \zeta_3^2+4200 \zeta_7  \right) + \nonumber\\
&\lambda^5 \left( 7092 \zeta_3+1980 \zeta_5+ 864 \zeta _3^2+2667 \zeta_7-7560 \zeta_3 \zeta_5-52920 \zeta_9 \right)\,.
\end{align}
This mismatch occurs when all bridges in the three-point function have length one. The triple wrapping $A_{\{1,1,1\}}$ was originally expected at six loops, but our results seem to indicate that it contributes already at five loops with 
\begin{equation}
\mathcal A_{(1,1,1)} = \lambda^5 \left(5832 \zeta_3 -16200 \zeta_5+32130 \zeta_7-14256 \zeta_3 \zeta_5-9072 \zeta_9\right) \,.
\end{equation}
This is not unexpected, as the two virtual excitations in the adjacent bridges make the original proposal for the triple wrapping divergent. We expect that the required regularization of this term, along the lines of \cite{Basso:2017muf}, will anticipate its contribution to five loops.

In order to test that the mismatch is indeed due to a triple wrapping, we also studied the OPE limit of the following correlators
\begin{equation}
\langle \mathcal{O}_2(x_1)\mathcal{O}_3(x_2)\mathcal{O}_{2+m}(x_3) \mathcal{O}_{3+m}(x_4)\rangle \,.
\end{equation}
We isolated the twist three contributions for all values of $m$ and showed that in this case the results are perfectly compatible with the contribution of opposed wrapping for all bridge lengths, proving in that way that the mismatch occurs only when all bridges have length one.

\section{Conclusions}\label{conc}

We have obtained all four-point functions of protected operators in $\mathcal{N}=4$ SYM up to the five-loop order. Our method relies on a combination of two techniques: first we consider light-cone OPE relations between integrands of different correlators, and then we take the euclidean OPE limit of the integrated four-point functions and compare with data obtained from integrability. We extract a myriad of OPE coefficients and check that they perfectly agree with OPE data obtained with integrability (which we did not have to use to fix the correlators).

While we have found convincing evidence that the saturation bound in the $R$-charge structures of four-point functions at $\ell$ loops is $(\ell-1)$, it would be interesting to prove this statement. Our results also seem to indicate that the strong version of the light-cone OPE relations is valid in $\mathcal N=4$ SYM. This fact should be examined in more detail, as a proof of its validity would tremendously simplify the study of four-point functions of protected operators at higher loops. 

By focusing on the correlator of four $\mathcal O_{20'}$ operators, we have shown that new wrapping effects appear in the hexagon approach to three-point functions at five loops. This is an example of a fruitful interplay between the integrability machinery and the more standard perturbative quantum field theory methods, and it would now be important to obtain this result from the integrability point of view. Since the regularization of hexagon form factors seems to anticipate wrapping corrections, one should study what are the implications on the positivity of the hexagon perturbation theory \cite{Eden:2018vug}.

It is also possible to employ integrability in the study of four-point functions, by using the method of hexagonalization. It would be interesting to evaluate the observables obtained in this work with such methods, as there is now a point of comparison. Furthermore, by picking specific polarizations for the external operators one can probe different finite-size corrections of the four-point functions. In principle, this could lead to integrability representations of higher-point conformal integrals, in the spirit of \cite{Basso:2017jwq}.

In this work we considered the euclidean OPE limit of the four-point functions, which was obtained at leading order with the method of asymptotic expansions. However, it would be extremely helpful to evaluate exactly all conformal integrals that appear in the correlators, since that would allow us to take other relevant limits which cannot be accessed by asymptotic expansions.

\section*{Acknowledgments}
We are grateful to Frank Coronado for many useful discussions.

V.G. is funded by FAPESP grant 2015/14796-7 and CERN/FIS-NUC/0045/2015. The work of A.G. is supported by the Knut and Alice Wallenberg Foundation under grant \#2015-0083. R.P. is supported by SFI grant 15/CDA/3472.

\appendix

\section{Asymptotic Expansions} \label{AppAE}

The integrals appearing in the four-point function at four and five loops are hard to compute for generic values of the cross ratios $u$ and $v$\footnote{The interested reader can find the most recent advances in the evaluation of conformal integrals in \cite{Eden:2016dir,Basso:2017jwq,Golz:2015rea}.}. However, for our purposes it is sufficient to extract the values of these integrals in the euclidean OPE limit $u\to 0$ and $v\to 1$, which can be done with the method of asymptotic expansions.
This method was  introduced in \cite{Beneke:1997zp,Eden} and has been applied recently to compute five-loop p-integrals and structure constants in $\mathcal N=4$ SYM \cite{Vasco,Fiveloops}.

The $\ell$-loop correlator depends on four external points $\{x_1,x_2,x_3,x_4 \}$ and $\ell$ internal points which we integrate over, and all propagators are differences of the form
\begin{equation}
x_{ij}=x_i-x_j \,.
\end{equation}
Conformal symmetry can be used to send $x_1$ to the origin and $x_4$ to infinity, and the final result is naturally expressed in terms of the ratios
\begin{equation}
u=\frac{x_2^2}{x_3^2}\,, \qquad v=\frac{x_{23}^2}{x_3^2}\,.
\end{equation} 
The structure of the four-point function is not arbitrary since the short-distance singularities are constrained by the OPE data of the theory. We are interested in the short-distance limit of the integrals, or in other words we want to study the behavior of the integral when $x_2$ approaches the origin.
The main idea behind the method of asymptotic expansions is to divide each integration domain in several regions, so that it is possible to take the short-distance limit inside the integral. In practice we divide the integration over each internal point $x_i$ in two different regions: one where the integration point is close to $x_2$ and one where it is close to $x_3$.
In each of these regions we can expand the propagators in the following way: 
\begin{equation}
\frac{1}{(x_2 - x_i)^2} 
= \sum_{n=0}^\infty \frac{(2 x_2 \cdot x_i - x_2^2)^n}{(x_i^2)^{n+1}} 
\quad (\text{if $x_2^2< x_i^2$})\,,
\label{eq:propagator-expansion1}%
\end{equation}
\begin{equation}
\frac{1}{(x_3 - x_i)^2} 
= \sum_{n=0}^\infty \frac{(2 x_3 \cdot x_i - x_i^2)^n}{(x_3^2)^{n+1}} 
\quad (\text{if $x_i^2< x_3^2$})\,.
\label{eq:propagator-expansion2}%
\end{equation}
There are $2^\ell$ regions corresponding to the $\ell$ integration points and in each of these regions the original integral is expressed as a product of two-point integrals. If $k$ integration variables are in the region close to $x_2$, then the $k$-loop  integral with external points $x_1$ and $x_2$ multiplies an  $(\ell-k)$-loop integral with external points $x_1$ and $x_3$.

Then we use the fact that integrals are not all independent since they satisfy IBP identities. In particular this makes it possible to express any two-point integral as a linear combination of master integrals. These identities can be obtained using a computer implementation of the Laporta algorithm such as FIRE \cite{Smirnov:2014hma}. The values of the master integrals used for this computation were evaluated in \cite{pintegrals}. 

The integrals used here might be useful for other studies and for this reason we include them in an auxiliary file. We have computed the four-loop integrals up to $u^0$ and $(1-v)^4$, while the expansions of the five-loop integrals are at $u^0$ and for $v=1$.

\bibliography{biblio}
\bibliographystyle{JHEP}

\end{document}